\documentclass[final,times,numbers,compress]{elsarticle}

\usepackage{amssymb}
\usepackage{amsmath}
\usepackage{graphicx}
\usepackage{subcaption}
\usepackage{xurl}      
\usepackage[hidelinks]{hyperref}

\begin{document}
\begin{frontmatter}

\title{Application and Optimization of the GET Electronics System for Double-Sided Silicon Strip Detector Readout\tnoteref{arxivnotice}}

\tnotetext[arxivnotice]{\textcopyright{} 2026. This manuscript version is made available under the
CC-BY-NC-ND 4.0 license:
\href{https://creativecommons.org/licenses/by-nc-nd/4.0/}{\nolinkurl{https://creativecommons.org/licenses/by-nc-nd/4.0/}}.}

\author[PKU]{Jiawei BIAN} 
\author[PKU]{Zaihong YANG\corref{cor1}} 
\ead{zaihong.yang@pku.edu.cn}
\author[PKU]{Qite LI} 
\author[PKU]{Zeyu DU} 
\author[USTC]{Xuan WANG}
\author[IMP]{Ningtao ZHANG}
\author[FDU]{Pengjie LI}
\author[USTC]{Zhongtao SHEN}
\author[PKU]{Kaijie ZHOU} 
\author[PKU]{Cheng WANG} 

\cortext[cor1]{Corresponding author}

\affiliation[PKU]{organization={School of Physics, State Key Laboratory of Nuclear Physics and Technology, Peking University},
            city={Beijing},
            postcode={100871}, 
            country={China}
            }
\affiliation[USTC]{organization={University of Science and Technology of China},
            city={Hefei},
            postcode={230026}, 
            country={China}
            }
\affiliation[IMP]{organization={Institute of Modern Physics},
            city={Lanzhou},
            postcode={730000}, 
            country={China}
            }
\affiliation[FDU]{organization={Fudan University},
            city={Shanghai},
            postcode={200433},
            country={China}
            }

\begin{abstract}
Double-sided silicon strip detectors (DSSDs) are widely used for charged-particle detection in nuclear reaction experiments.
However, the large number of readout channels associated with DSSDs places stringent demands on front-end electronics and data acquisition systems, particularly in experiments employing large detector arrays.
In this work, we investigate the application and optimization of the Generic Electronics for Time Projection Chambers (GET) system for multi-channel DSSD readout.
Through optimization of the DC-blocking capacitors on the custom-designed interface board and waveform-based energy extraction algorithms,
the readout system achieves an energy resolution of $\sigma$ = 22 keV for 5.49 MeV alpha particles.
Its performance is further validated in a representative nuclear physics experiment employing charged-particle telescopes. 
The results demonstrate that the GET system provides a versatile and scalable waveform-sampling solution for multi-channel DSSD readout.
\end{abstract}




\end{frontmatter}

\section{Introduction}\label{sec1}
Nuclear clustering, where nucleons (protons and neutrons) form strongly correlated substructures such as $\alpha$ clusters, 
represents one of the most prominent and widely studied emergent phenomena in nuclear systems\cite{wei-clustering-2024,ye-physics-2025,freer-microscopic-2018}.
Extensive experimental and theoretical investigations over the past decades have established clustering as a general feature of nuclear many-body dynamics, 
manifesting across a broad range of nuclear systems from light stable nuclei to neutron-rich and heavy nuclei\cite{yang-observation-2014,tanaka-formation-2021,li-validation-2023}.
Studies of nuclear clustering are essential for the development of theories describing finite nuclei, 
and also provide key insights into the properties of nuclear matter, 
with far-reaching implications for the nuclear equation of state and the physics of neutron stars\cite{yang-clustering-2021, oertel-equations-2017}.

From an experimental perspective, 
investigations of clustering---as well as many other topics in nuclear structure studies, 
including nuclear spectroscopy and transfer or knockout reaction measurements---rely critically on the precision measurements of energy, position, and emission angle of the charged-particle reaction products. 
Consequently, the development of high-resolution charged-particle detection systems remains a central task in experimental nuclear physics. 
In recent years, increasing emphasis has been placed on experiments with unstable nuclei using radioactive ion beams. 
Owing to the typically low beam intensities and small reaction cross sections, such experiments generally rely on large detector arrays to maximize the detection efficiency and thus the reaction yield, 
leading to increasingly stringent requirements in terms of energy resolution, granularity, scalability, and readout electronics.

Silicon detector arrays, often combined with scintillators to form $\mathrm{\Delta E-E}$ charged-particle telescopes, 
are widely used in nuclear physics experiments due to their excellent energy resolution, fast response, and radiation hardness. 
Over the past decades, a variety of charged-particle telescope arrays based on silicon detectors and scintillators have been developed worldwide, including 
TOGAXSI\cite{tanaka-designing-2023}, LASSA\cite{davin-lassa-2001}, MUST2\cite{pollacco-must2-2005}, HiRA\cite{wallace-high-2007}, and FARCOS\cite{acosta-farcos-2012,quattrocchi-farcos-2014,pagano-farcos-2015}.
In particular, DSSDs provide simultaneous energy and two-dimensional position information, 
making them well suited for experiments requiring high angular and energy resolution as well as large solid-angle coverage. 
However, the large number of readout channels associated with DSSDs poses stringent challenges for front-end electronics and data acquisition systems.

In parallel with advances in detector technology, waveform-sampling readout electronics have been developed in nuclear physics instrumentation, 
with the Generic Electronics for Time Projection Chambers (GET) system\cite{pollacco-get-2018,pollacco-get-2012} being a prominent example. 
By digitizing the full detector signal waveform, such systems enable flexible offline signal processing and improved noise rejection. 
The GET system, originally developed for large-scale TPC applications, is a modular, high-density readout system capable of handling thousands of channels with configurable gain, shaping, and sampling parameters. 
While GET has been widely deployed for gas detectors\cite{shane-srit-2015, bradt-commissioning-2017, mauss-commissioning-2019}, its application to silicon detector readout---particularly for DSSDs---has not yet been well established.

In this work, we investigate the application and optimization of the GET electronics system for the readout of DSSDs used in charged-particle detection. 
Using custom-designed interface boards, DSSD signals are coupled to the GET front-end electronics. 
We optimize the DC-blocking capacitors and waveform-based energy extraction algorithms, aiming at improving energy resolution and robustness against electronic noise. 
The optimized system achieves an energy resolution of $\sigma$=22 keV for 5.49 MeV alpha particles 
and has been validated in a representative nuclear physics experiment employing charged-particle telescopes. 
The results demonstrate that the GET system provides a versatile and scalable solution for multi-channel DSSD readout. 

\section{GET system and test setup}\label{sec2}

The GET system was originally developed through a collaboration between CEA/\allowbreak IRFU, CENBG, GANIL, RIKEN, and NSCL/MSU. 
It is characterized by high channel density and waveform-sampling readout operated at relatively low trigger rates (typically $\sim$ 100 Hz),
making it well suited for nuclear physics experiments involving detectors with a large number of readout channels. 

The GET architecture is based on the AGET (ASIC for GET) chip, which provides waveform sampling and signal processing.
Each AGET is capable of handling 64 signal channels together with 4 Fixed Pattern Noise (FPN) channels.
It features a 12-bit resolution and a maximum sampling frequency of 100 MHz, enabling precise digitization of typical nuclear detector signals.
Four AGET chips are integrated with a 4-channel analog-to-digital converter (ADC), a field-programmable gate array (FPGA), and supporting circuitry to form a front-end board known as AsAd (ASIC and ADC) board.
Each AsAd board therefore processes 256 signal channels.
Up to four AsAd boards can be controlled by a single Concentration Board (CoBo),  
and multiple CoBos can be further assembled into a standard Micro Telecommunications Computing Architecture (MicroTCA) crate.
This hierarchical architecture based on ASIC-level integration enables GET to scale efficiently to thousands of readout channels within a single crate.

\begin{figure}[htbp]
  \centering
  \includegraphics[width=0.7\textwidth]{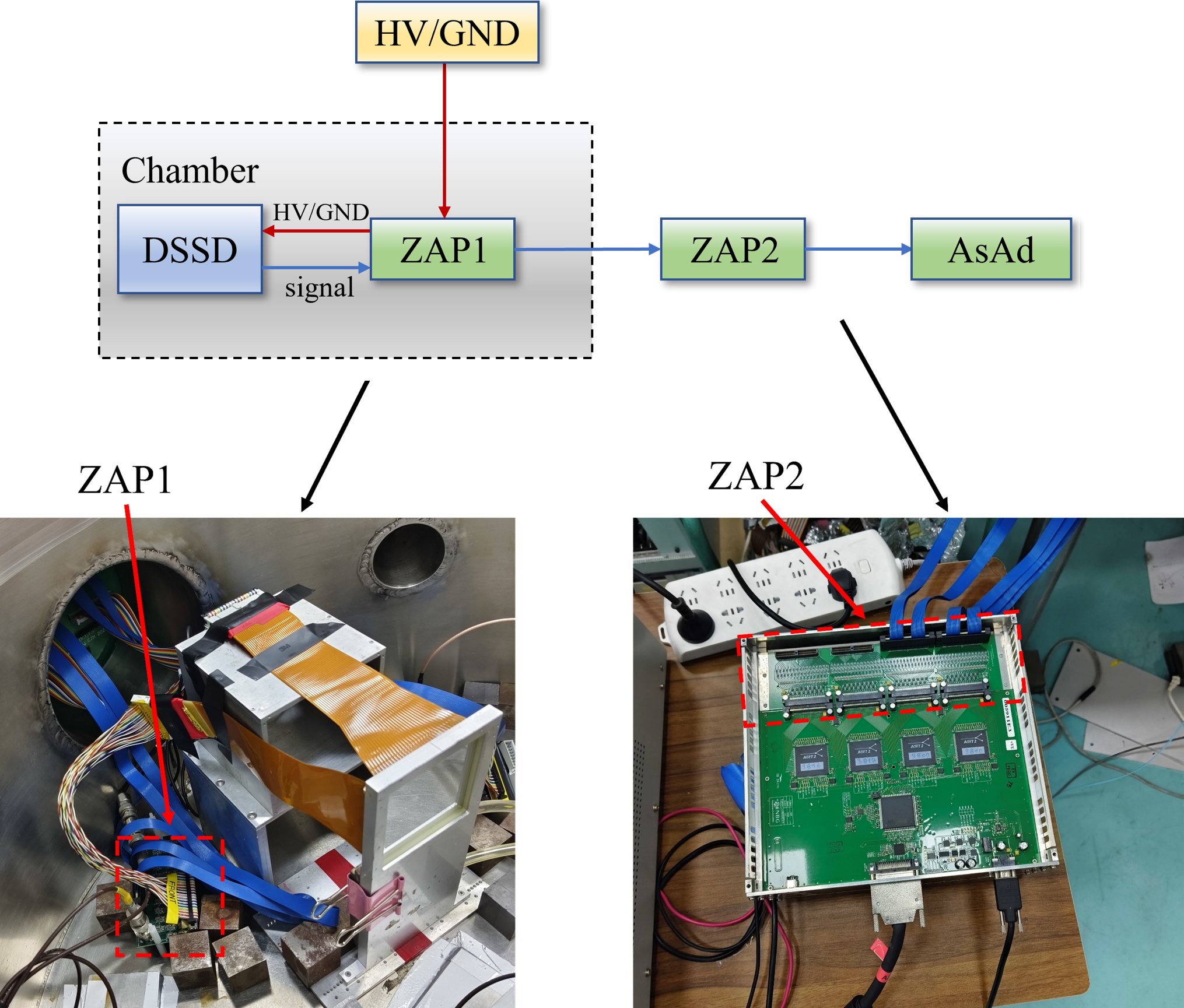}
\caption{Schematic diagram and photograph of the offline test setup for DSSD readout using the GET system. The $^{241}$Am source was placed  in front of the DSSD (not shown in figure).}\label{fig-test_setup}
\end{figure}

Before applying the GET system to the signal readout of DSSDs in physics experiments with radioactive ion beams, 
a dedicated offline test platform was established to verify the feasibility of DSSD signal readout and to optimize the system performance.
The test setup consisted of a single DSSD connected to one AsAd board.
Measurements were performed using a $^{241}$Am $\mathrm{\alpha}$-particle source and a three-$\mathrm{\alpha}$ source. 
A schematic view of the test setup is shown in Fig. \ref{fig-test_setup}.

The DSSD used in this study was a  BB7 detector manufactured by Micron Semiconductor Ltd, with a thickness of 1000 $\mathrm{\mu}$m.
This thickness is sufficient to fully stop $\mathrm{\alpha}$ particles with energies around 5 MeV.
The detector has a typical capacitance of about 20 pF per channel\cite{Micron2009Catalogue}
During the experiments, the pressure in the target chamber was maintained at approximately 100 Pa.
Signal transmission from the detector to the AsAd board was carried out using EQCD-040-12.00-SED-SEU-1 cables from Samtec LLC (hereafter referred to as Samtec cables).
Owing to their coaxial design, these cables effectively suppress external electromagnetic interference (EMI) during signal transmission,
which is beneficial for large detector arrays where longer signal cables may be required and flexible detector-electronics layouts are desirable.
Two types of custom-designed interface boards, referred to as ZAP1 and ZAP2, were employed at both sides of DSSD output and at the AsAd board input. 
The ZAP1 board provides several essential functions, including high-voltage bias supply for the junction side of the DSSD, GND for ohmic side of the DSSD, DC blocking, common grounding, and conversion from the DSSD output connectors to Samtec cable interfaces.
The ZAP2 board further adapts the signals from the Samtec cable inputs to the AsAd board. 
In the offline tests, the GET system was operated in multiplicity-trigger mode, 
in which a self-trigger signal was generated when the number of threshold-crossing channels exceeded a preset multiplicity. 
This mode simplified the trigger logic for the DSSD tests.

DC-blocking capacitors are mounted on both ZAP boards and are connected in series in the signal path. 
The capacitor on the ZAP2 board has a fixed capacitance of 220 pF, which blocks the DC component while limiting the signal amplitude delivered to the AsAd input. 
In contrast, the capacitor on the ZAP1 board is variable, allowing systematic investigation of the effects of different capacitance values on signal characteristics and energy resolution. 
Optimization of this capacitance is one of the main objectives of the offline test.

\section{Optimization of DC-blocking capacitor and offline analysis algorithms}\label{sec3}

\subsection{DC-blocking capacitor optimization and effects of peaking time} \label{sec3-1}

\begin{figure}[htbp]
  \centering
  \includegraphics[width=0.7\textwidth]{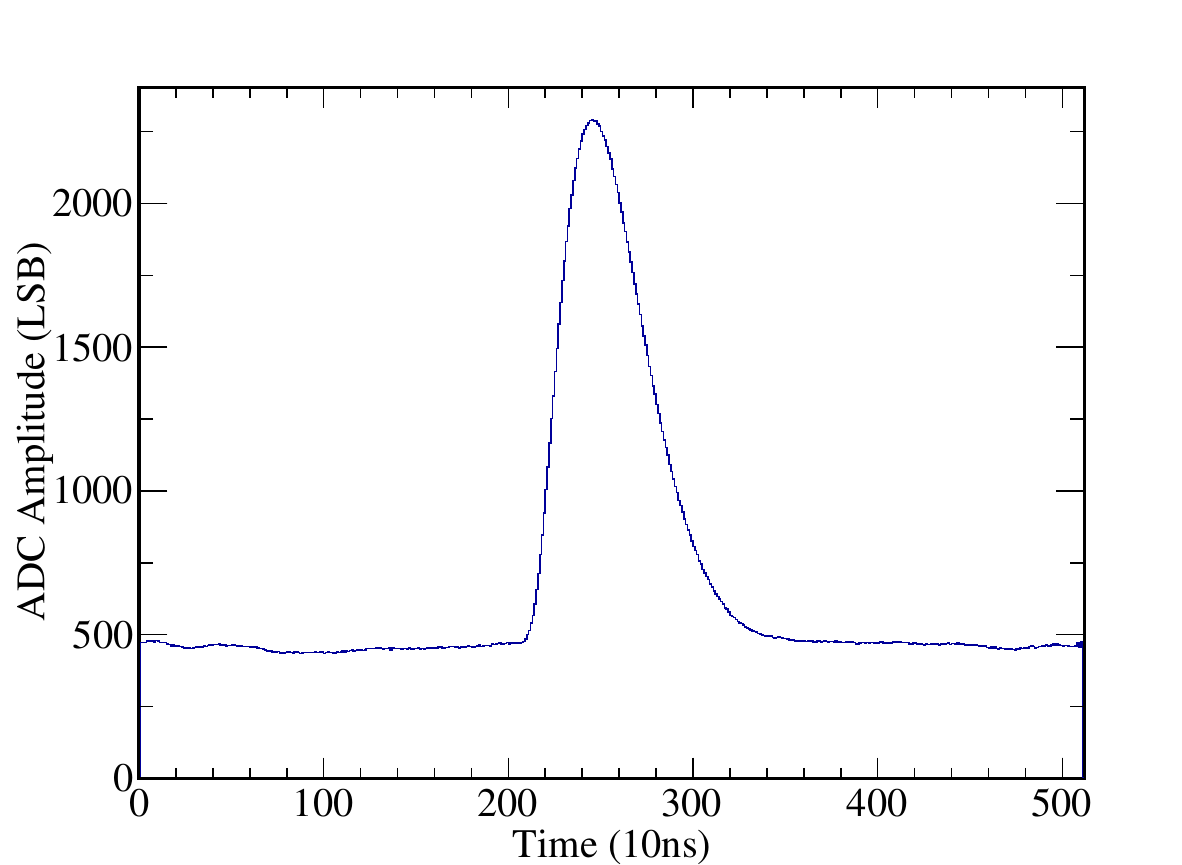}
  \caption{A representative digitized signal waveform for a 5.49-MeV $\mathrm{\alpha}$ particle, recorded by the GET system with a peaking time of 334 ns .}\label{fig-dig_waveform} 
\end{figure}

Tests for DC-blocking capacitor optimization were conducted using a $^{241}$Am radioactive source, 
which emits $\mathrm{\alpha}$ particles with an energy of around 5.49 MeV.
The $\mathrm{\alpha}$ source was collimated and positioned in front of the center of the DSSD, 
By adjusting the distance between the source and the detector, the trigger rate was controlled within the range of 50--100 Hz, 
ensuring the GET system operated under a relatively low load.
A representative digitized signal waveform is shown in Fig. \ref{fig-dig_waveform}.

It should be noted that the energy deposition of $\mathrm{\alpha}$ particles in the DSSD generates a charge of approximately 240 fC. 
Given that the GET system provides only four discrete gain settings---120 fC, 240 fC, 1 pC, and 10 pC---the gain was set to 240 fC during the test.
The DC-blocking capacitors were optimized accordingly (see below) to adjust the signal amplitude to an appropriate level within the available dynamic range of the GET system, 
thereby ensuring neither signal saturation nor insufficient utilization of the ADC resolution. 
A similar strategy was adopted in the physics measurement experiments, 
where the 1 pC gain setting was used to accommodate higher deposited energies; 
this will be discussed in Section \ref{sec4}.

\begin{figure}[htbp]
  \centering
  \includegraphics[width=0.7\textwidth]{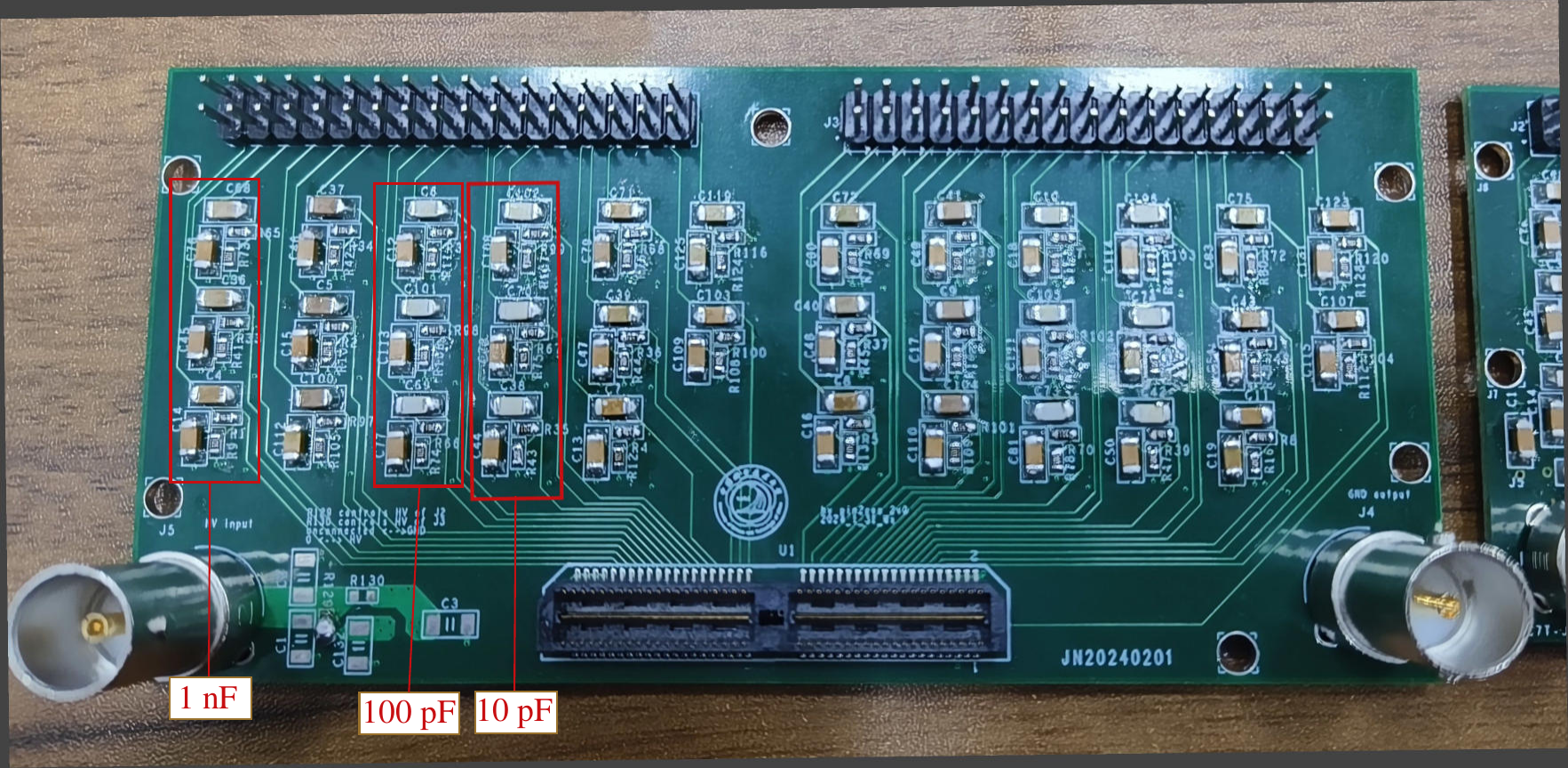}
  \caption{ZAP1-T interface board configured with different DC-blocking capacitance values on individual channels.}\label{fig-multic_zap}
\end{figure}

To systematically investigate the influence of the DC-blocking capacitance on signal amplitude and energy resolution, 
a dedicated ZAP1-T board was designed with varying capacitance values implemented on different channels. 
The ZAP1-T board was configured with capacitors of four nominal capacitance values (10 pF, 100 pF, 1 nF, and 10 nF) on different channels. 
When combined in series with the DC-blocking capacitor (fixed at 220 pF ) on the ZAP2 board, 
these values correspond to effective total DC-blocking capacitances of 9.56 pF, 68.8 pF, 180 pF, and 215 pF, respectively.
As shown in Fig. \ref{fig-multic_zap}, three channels were assigned to each of the first three capacitance values, while the remaining channels were equipped with 10 nF capacitors. 

Signal amplitudes and energy resolutions were analyzed for all channels. The data processing procedure was as follows:
The baseline and background contributions measured by the FPN channels were first subtracted from the raw waveforms for each channel. 
The residual baseline offset was then corrected using the mean value of the first 50 sampling points\cite{giovinazzo-get-2016,bian-testing-2025}.
\begin{equation}
  \begin{aligned}
  &s_i[t] = s_{i,\text{raw}}[t] - \text{fpn}_{\text{ave}}[t]\\  
  &s_{i,c}[t] = s_i[t] - \frac{\sum_{j=1}^{50}s_i[j]}{50},
  \end{aligned}
\end{equation}
where $s_{i,\text{raw}}[t]$ denotes the amplitude of channel $i$ at the $t$-th sampling point.
$\text{fpn}_{\text{ave}}$ represents the average of all FPN channels, and $s_{i,c}$ is the corrected waveform.

For the present analysis, 
the peak value of the corrected waveform  was used as a measure of the signal amplitude and thus the deposited energy (below referred to as ``Direct peak-finding''). 
The measured energy spectra were then constructed for each channel, 
and the energy resolution was extracted through Gaussian fitting, 
defined as the ratio of the full width at half maximum (FWHM) to the peak energy.

\begin{figure*}[htbp]
  \centering
  \includegraphics[width=1\textwidth]{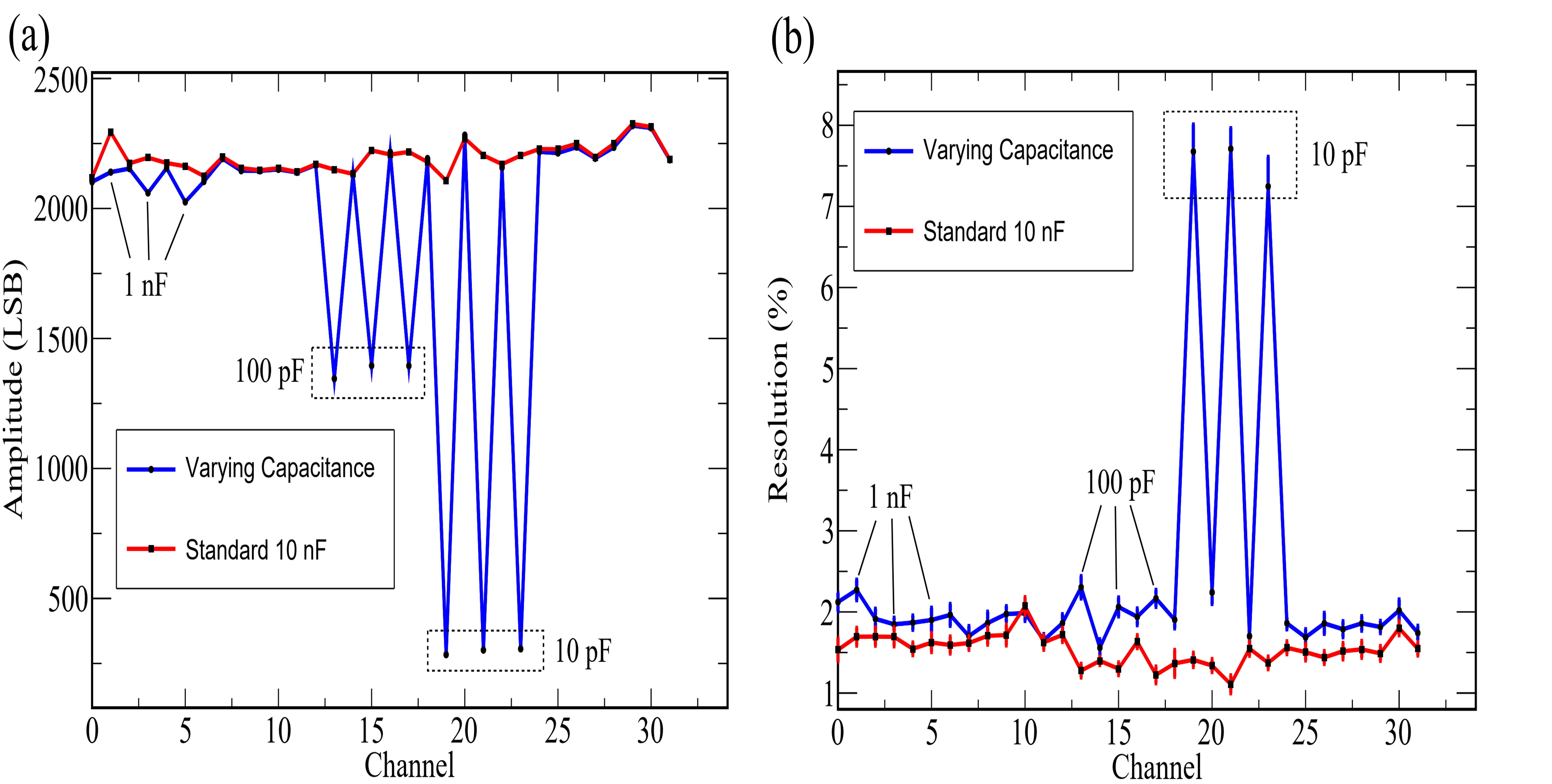}
\caption{(a) Signal amplitude and (b) energy resolution obtained for different DC-blocking capacitance values. Results from a standard ZAP1 board equipped with 10-nF capacitors are shown for comparison.}\label{fig-multiC_results}
\end{figure*}

Fig. \ref{fig-multiC_results} presents the measured signal amplitude and energy resolution for different channels,
with the corresponding DC-blocking capacitance values indicated.
For comparison, a standard ZAP1 board fully equipped with 10 nF DC-blocking capacitors was also tested under identical conditions.
It can be observed that for capacitances less than 220 pF, 
both the signal magnitude and the energy resolution improve with increasing DC-blocking capacitance values.
Up to the maximum blocking capacitance investigated in this work, 
the resolution continues to improve with increasing capacitance. 
Further studies with larger effective capacitance values would be useful to determine whether the resolution eventually reaches an optimum beyond the dynamic-range limitations considered here. 
However, given that the GET system has a 12-bit sampling resolution, with a typical baseline level ranging from 300 to 500 LSB, 
the effective dynamic range is approximately 3600 LSB. 
Fig. 4(a) shows that with a DC-blocking capacitance of 10 nF the signal amplitude matches best the available dynamic range of the GET system. 
Therefore, 10 nF was selected as the optimal DC-blocking capacitance for the ZAP1 board.

\begin{figure}[htbp]
  \centering
  \includegraphics[width=0.7\textwidth]{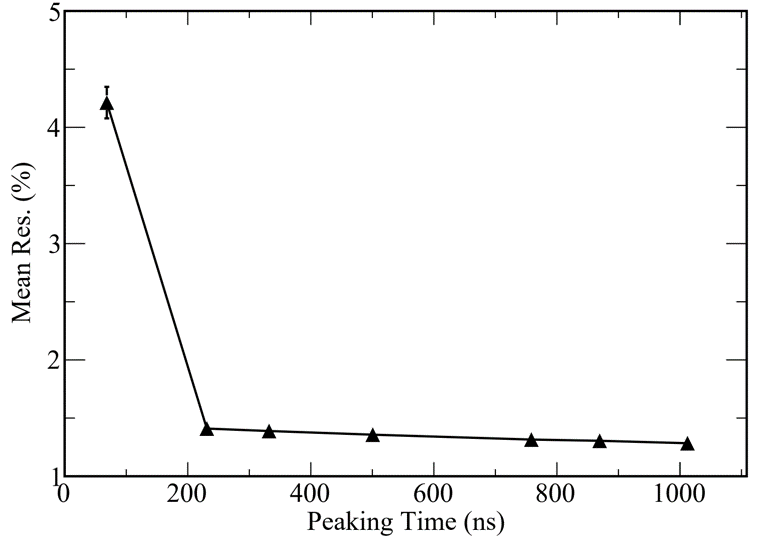}
\caption{Mean resolution as a function of the peaking time.}\label{fig-peakingtime}
\end{figure}

The effect of peaking time on the energy resolution was investigated using the optimized electronics configuration. 
With all ZAP-1 DC-blocking capacitors set to 10 nF, the energy resolution of each channel was measured at peaking times ranging from 70 ns to 1014 ns. 
The dependence of the average energy resolution on the peaking time is shown in Figure \ref{fig-peakingtime}. 
The resolution is significantly worse at the shortest peaking time of 70 ns than at longer peaking times. 
This behavior can be attributed to the limitation of sampling rate. For very short waveforms, 
a sampling frequency of 100 MHz leads to insufficient sampling points around the peak, thereby reducing the accuracy of amplitude extraction.
In addition, short waveforms are more susceptible to high-frequency noise. 
Both effects are effectively suppressed as the waveform width increases, leading to a gradual improvement in energy resolution. 
The noise robustness of the energy-extraction algorithms is further discussed in Section \ref{sec3-2}.

In this work, the peaking time was set to 334 ns, as a compromise between resolution and trigger rate. 
It should be noted that, a longer peaking time may lead to better energy resolution if the trigger rate and data-transmission bandwidth are not limiting factors.
For the AGET chip used in this work, the maximum available peaking time of 1014 ns would thus give the best resolution. 
The next-generation AGET chip, STAGE (SEcond stAGe of the AGET), provides peaking times up to 8 $\mathrm{\mu s}$\cite{Aumann2022,Baron2019,Wang2025}.
Therefore, further improvement in energy resolution may be expected using STAGE with longer peaking times.

\subsection{Investigation of energy extraction algorithms}\label{sec3-2}

Nuclear reaction experiments conducted at beamlines of accelerator facilities operate in complex electromagnetic environments,
where detector signals are affected by various electronic noise sources, including white noise, baseline fluctuations, and EMI from beamline infrastructure and auxiliary electronics. 
These noise contributions distort the recorded waveforms, particularly the baseline, and introduce uncertainties in signal amplitude determination, 
leading to degraded energy resolution. 
Taking advantage of the full temporal information provided by waveform-sampling readout systems such as GET, 
properly optimized waveform-based energy extraction algorithms have the potential to suppress the effects of noise and baseline variations, thereby enhancing energy resolution. 
Such software-based approaches complement hardware-based noise reduction methods including shielding and grounding.

To investigate waveform-based energy extraction methods,
an $\mathrm{\alpha}$-particle source consisting of three components---$^{239}$Pu, $^{241}$Am and $^{244}$Cm---was used for calibration and performance evaluation. 
These isotopes emit $\mathrm{\alpha}$ particles with energies of 5.15 MeV, 5.49 MeV, and 5.80 MeV, respectively.
During the measurements, shielding and grounding were applied to all signal cables and electronic units in the experimental setup to minimize externally introduced noise.

Two waveform fitting algorithms were investigated and are described below.

\noindent \textbf{Quadratic function fitting}: In this method, the sampled signal waveform in the vicinity of its maximum is approximated using a second-order Taylor expansion. 
Within a sufficiently narrow fitting window, the waveform exhibits characteristic quadratic behavior, allowing the signal amplitude to be extracted through parabolic fitting. 
This approach is computationally efficient and relies only on the local waveform information around the peak region.

\begin{figure}[htbp]
  \centering
  \includegraphics[width=0.7\textwidth]{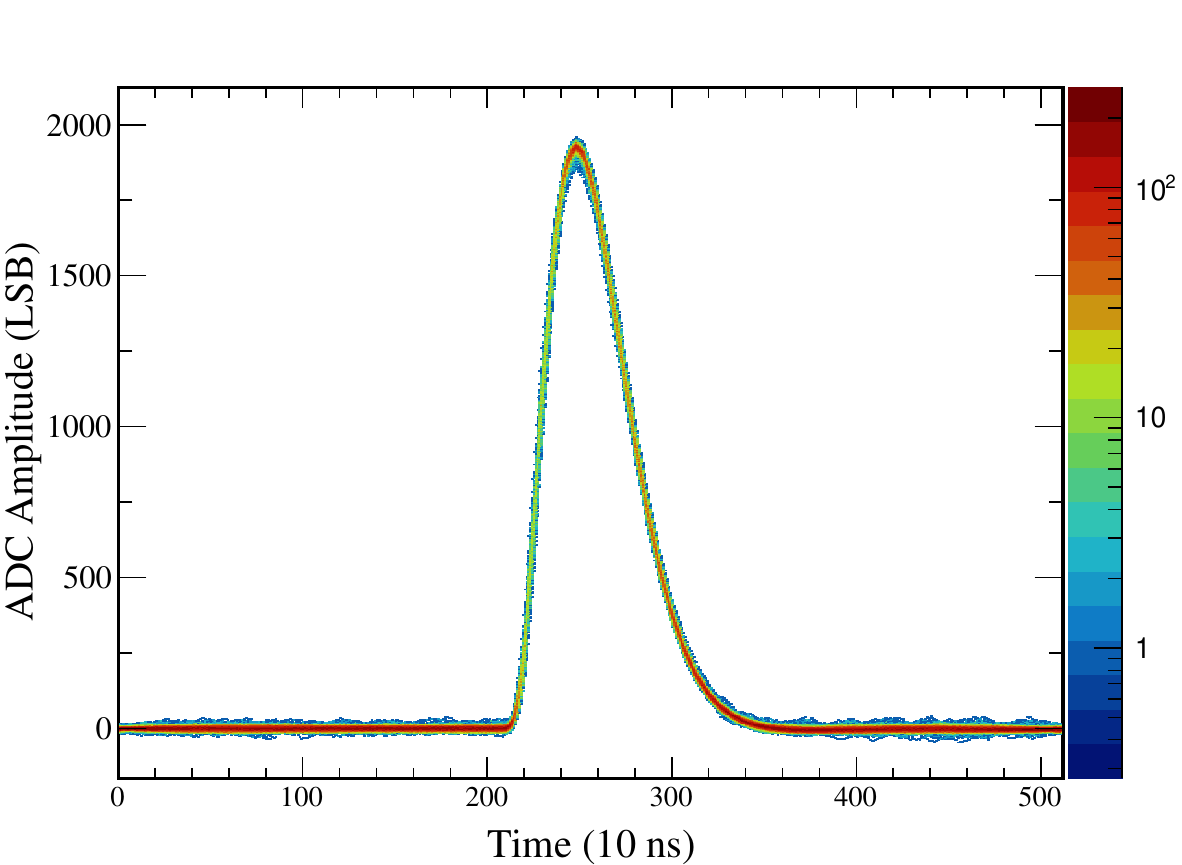}
\caption{Superimposed and averaged signal waveforms for a representative channel, used to construct the template waveform for template waveform fitting.}\label{fig-pilewave}
\end{figure}

\noindent \textbf{Template waveform fitting}: In principle, compared to quadratic function fitting, the precision and stability in extracting the signal magnitude 
and thus the energy resolution can be improved by fitting a larger portion of the waveform---or the full waveform---with a proper analytical function. 
However, due to the complex signal coupling between the DSSD and the GET electronics, as well as potential waveform differences among different incident particles, 
analytical functions may not adequately reproduce the actual signal shapes. 
To address this limitation, a template waveform fitting method was developed.
As shown in Fig. \ref{fig-pilewave}, signal waveforms corresponding to a single particle type and a limited energy range (5.49-MeV $\mathrm{\alpha}$ particles in the present case) were superimposed, 
and the average value at each sampling point was calculated to generate a reference waveform (template). 
This template waveform was then used to fit the signal waveforms of the $\mathrm{\alpha}$-source data and extract their amplitudes. 
To ensure consistency, in practical applications the template fitting should be restricted to waveforms corresponding to the same particle type and recorded under identical GET shaping parameter settings.

\begin{figure}[htbp]
  \centering
  \includegraphics[width=0.7\textwidth]{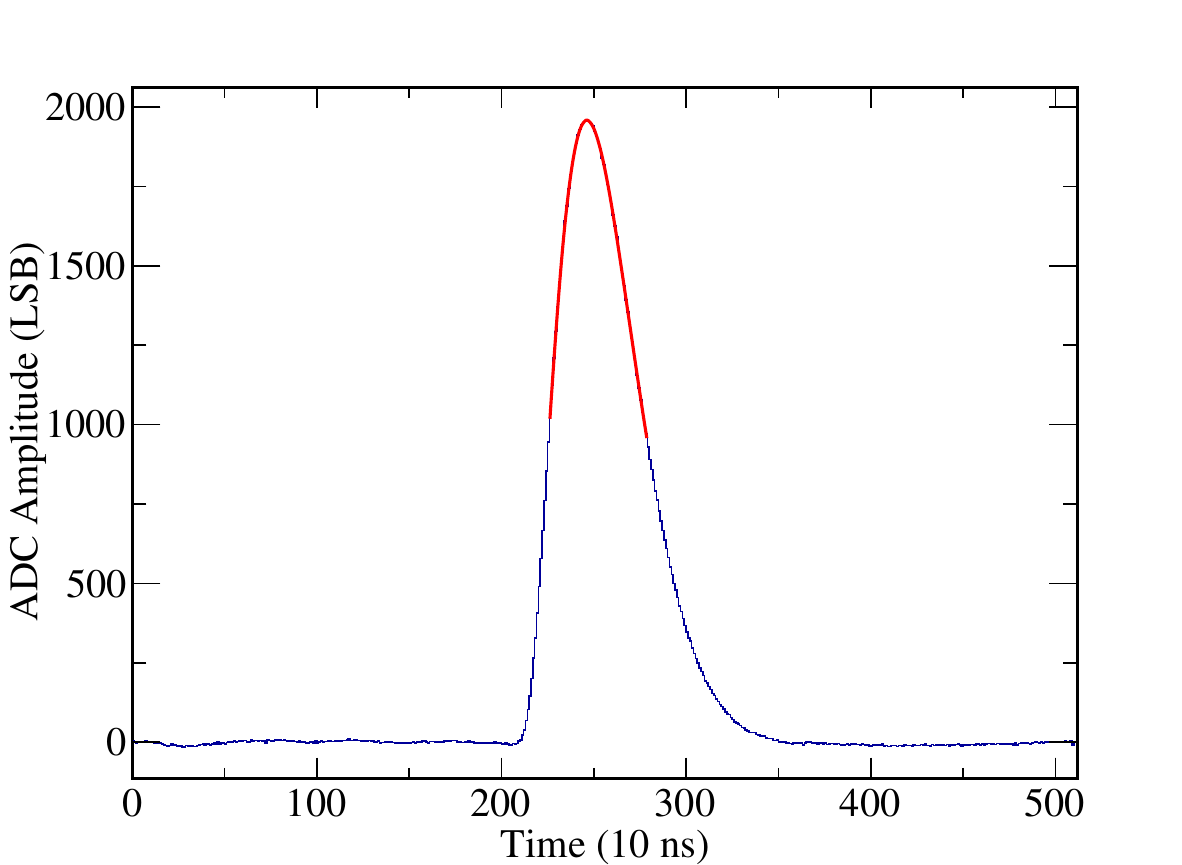}
\caption{Example of a baseline-corrected digitized waveform. The fitting range is defined by a relative amplitude threshold of $\eta$ = 50\%, and the red curve shows the result of the template waveform fitting.}\label{fig-fit_sample}
\end{figure}

For both fitting methods, the optimal fitting range was optimized through systematic scanning. 
A parameter, $\eta$, relative amplitude threshold with respect to the peak amplitude (maximum) of the waveform, 
was introduced to define the fitting range during the parameter scanning.
Only sampling points with relative amplitudes exceeding $\eta$ were included in the fit. 
An example of template waveform fitting with $\eta$=50\% is shown in Fig. \ref{fig-fit_sample}. 

\begin{figure}[htbp]
  \centering
  \includegraphics[width=0.7\textwidth]{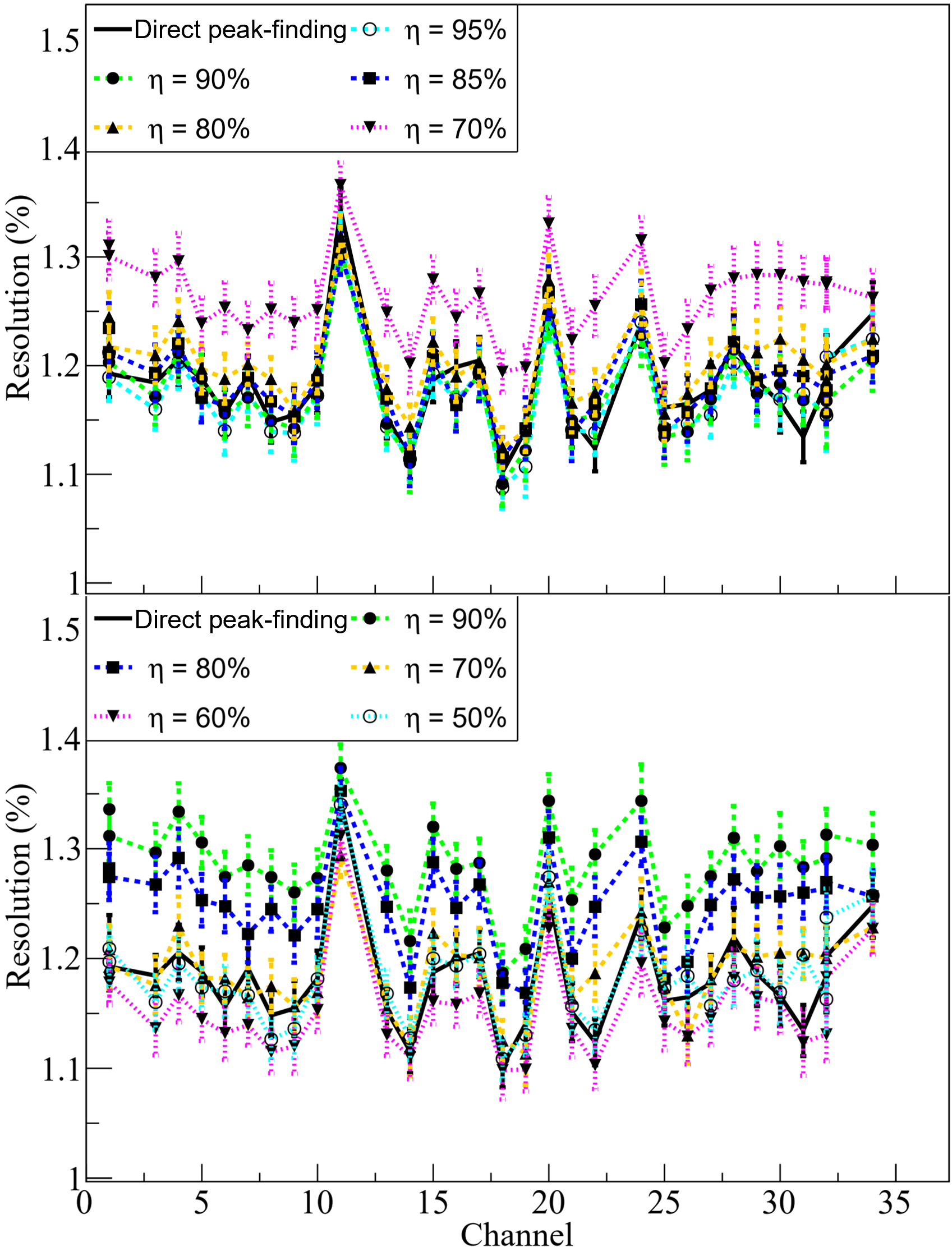}
\caption{Energy resolution of different channels for (upper panel) quadratic function fitting and (lower panel) template waveform fitting. Results obtained with different fitting ranges are shown for comparison.}\label{fig-res_fit_range}
\end{figure}

A comparison of the energy resolutions for 5.49-MeV $\mathrm{\alpha}$ particles obtained using direct peak-finding, quadratic function fitting, and template waveform fitting is presented in Fig. \ref{fig-res_fit_range} as a function of the fitting range. 
As the fitting range expands, the energy resolution achieved with template waveform fitting gradually improves, and reaches best results at around $\eta = 60 \%$.
The non-monotonic behavior around $\eta$= 60\% may result from the competition among the signal-to-noise ratio of the included waveform samples, the sensitivity to timing jitter and the reduction of statistical noise by multi-point fitting.
whereas the performance of quadratic function fitting deteriorates due to the inclusion of waveform regions that deviate from quadratic behavior. 
At their respective optimal fitting ranges, both fitting methods yield comparable energy resolutions. 
The resolution obtained from the direct peak-finding method here is better than that obtained in the  DC-blocking capacitor test (red curve in Fig. \ref{fig-multiC_results}(b)).
The most significant improvement comes from the energy calibration. 
Before calibration, the peak positions of all channels exhibited an offset of approximately -0.5 MeV,
which  led to an overestimated relative energy resolution.  
In addition, compared with the measurements presented in Section \ref{sec3-1}, 
the optimized shielding and grounding arrangements may also contribute to the performance improvement.

\begin{figure}[htbp]
  \centering
  \includegraphics[width=0.8\textwidth]{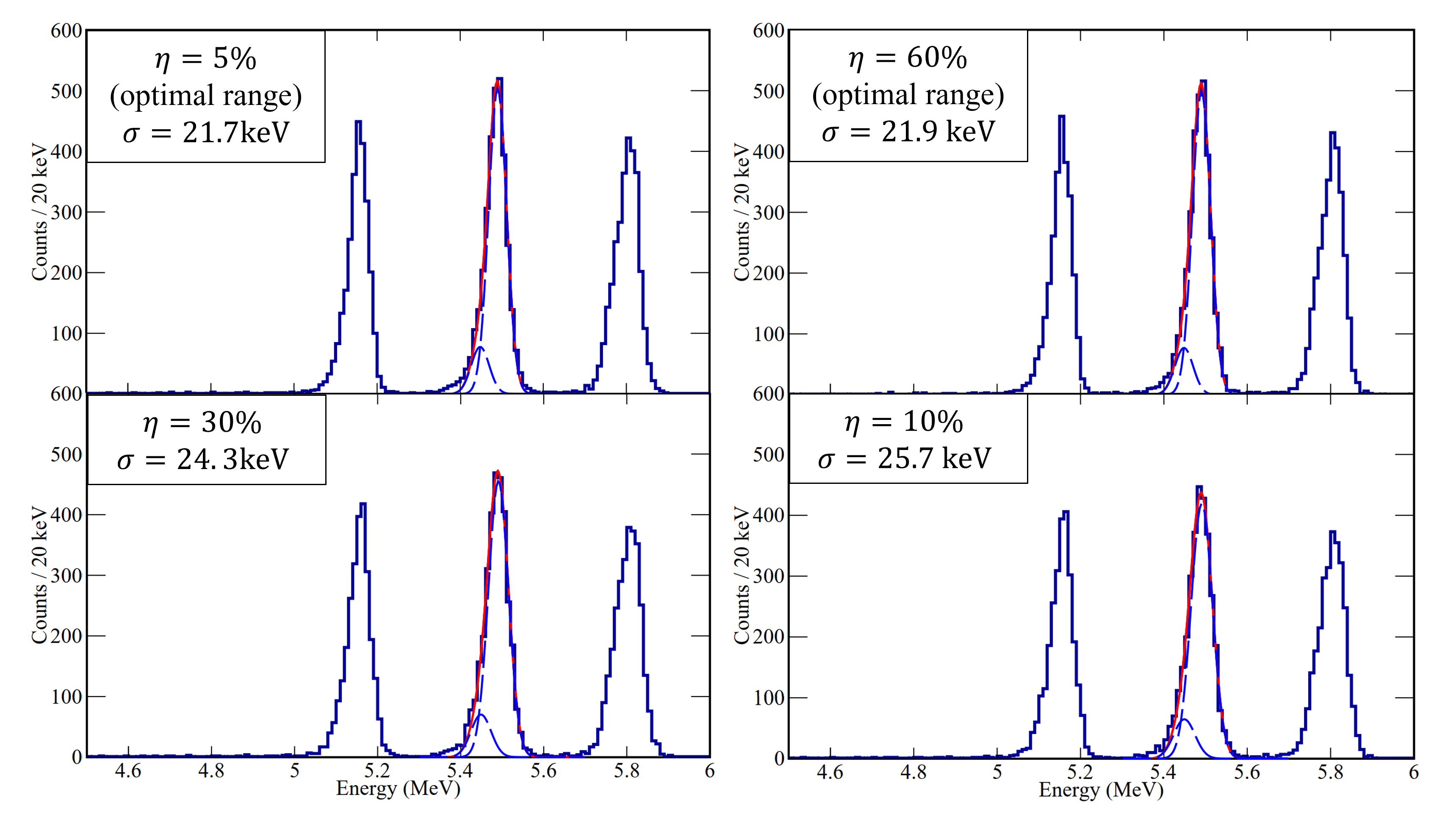}
\caption{Energy spectra for a representative channel (channel \#5) obtained using (a) quadratic function fitting and (b) template waveform fitting. Results corresponding to the optimal (upper panels) and non-optimal (lower panels) fitting ranges are shown for comparison. The red curve corresponds to double Gaussian function fit, and the blue dashed curves corresponds to the 5.44 MeV and 5.49 MeV components respectively.}\label{fig-energy_spec_range}
\end{figure}

Energy spectra obtained for a representative channel using the two fitting methods under both optimal and non-optimal fitting ranges are shown in Fig. \ref{fig-energy_spec_range}. 
A double Gaussian function was adopted to fit the peak, taking into account the two $\alpha$-particle branches of $^{241}$Am at 5.44 MeV and 5.49 MeV. 
The relative intensities of the two Gaussian components were fixed according to their decay branching ratio.
Under optimal conditions, an energy resolution of $\sigma$=22 keV is achieved for both fitting approaches. 
For comparison, the manufacturer's test data sheet provided by Micron Semiconductor Ltd. reports a typical measured energy resolution of FWHM = 60-100 keV at 5.49 MeV for this type of DSSD, corresponding to $\sigma = 25.5$-$42.5$ keV.
The resolution achieved with the optimized GET-based readout system in this work is slightly better than the typical values reported in the manufacturer’s datasheet.

The robustness of the three energy extraction methods against noise was evaluated using simulations.
Two types of noise were considered and added to the baseline-corrected waveforms: 
switching power supply noise and white noise. The switching power supply noise was modeled as
\begin{equation}
  n[i] = A\sin(2 \pi f \cdot t)\cdot \exp(-\frac{(t-t_0)^2}{2\sigma_s^2}), \ t = i/100 \text{ MHz}.
\end{equation}
where $A$ controls the amplitude, $f = 10 \text{MHz}$ denotes the center frequency, 
$\sigma_s = 300 \text{ ns}$ defines the envelope width, and $t_0$ is the center time of the envelope. 
The parameter $t_0$ was randomly sampled for each waveform.
White noise was simulated using Gaussian distribution with a standard deviation $\sigma$. 
The noise amplitudes $A$ and $\sigma$ were varied independently to simulate different noise conditions, 
and the corresponding average energy resolution across all channels was calculated for each extraction method. 
The fitting range for both fitting methods was fixed at their optimal values ($\eta = 96 \%$ and $\eta = 60 \%$, respectively). 

\begin{figure}[htbp]
  \centering
  \includegraphics[width=0.7\textwidth]{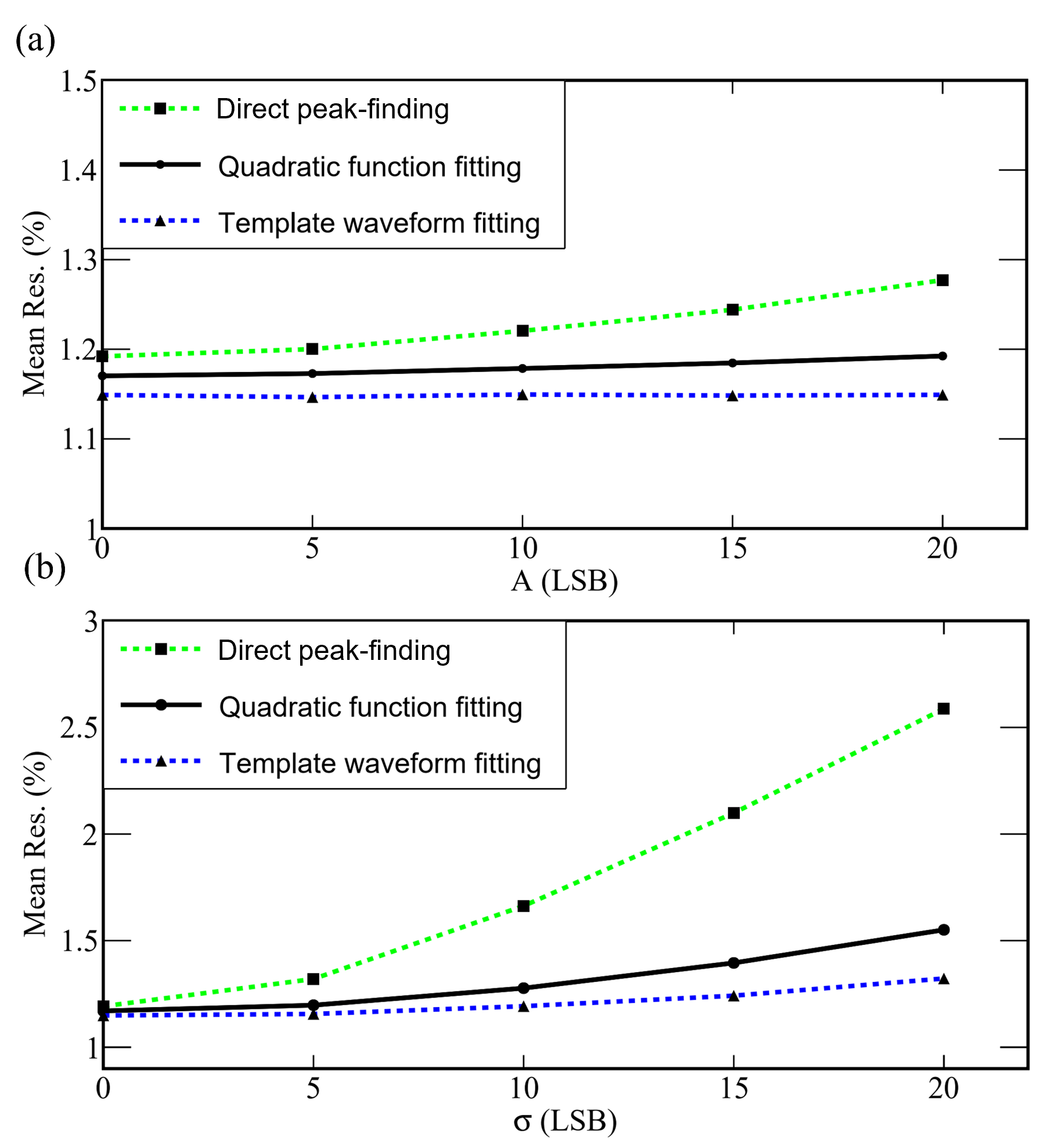}
\caption{Mean energy resolution as a function of (a) switching power-supply noise amplitude and (b) white-noise amplitude for the three energy extraction methods.}\label{fig-res_noise}
\end{figure}

Fig. \ref{fig-res_noise} presents a comparison of the energy resolutions obtained with the three methods as a function of noise amplitude.
As the noise level increases, the performance of direct peak-finding deteriorates rapidly. 
Quadratic function fitting exhibits improved robustness against noise, 
while template waveform fitting provides the best overall performance. 
In particular, template waveform fitting shows strong robustness against switching power supply noise.

Based on the above results, 
template waveform fitting yields the best resolution performance, 
with its advantage becoming increasingly pronounced  in high-noise environments.
Nevertheless, for applications requiring rapid data processing such as online monitoring during experiments, 
quadratic function fitting remains attractive due to  its simplicity and low computational cost.

\section{Application in a radioactive ion beam (RIB) experiment}\label{sec4}

To validate the GET-based readout system under realistic experimental conditions, 
a charged-particle measurement was carried out at the Radioactive Ion Beam Line in Lanzhou (RIBLL1) at the Heavy Ion Research Facility in Lanzhou (HIRFL), Institute of Modern Physics (IMP)\cite{sun-ribll-2003}. 
The experiment investigated breakup and cluster knockout reactions of unstable beryllium and carbon isotopes such as $^{10}$Be and $^{10}$C in inverse kinematics 
and served as a representative test case for the application of the GET system to DSSD-based telescope arrays in nuclear reaction studies.

\begin{figure*}[htbp]
  \centering
  \includegraphics[width=1\textwidth]{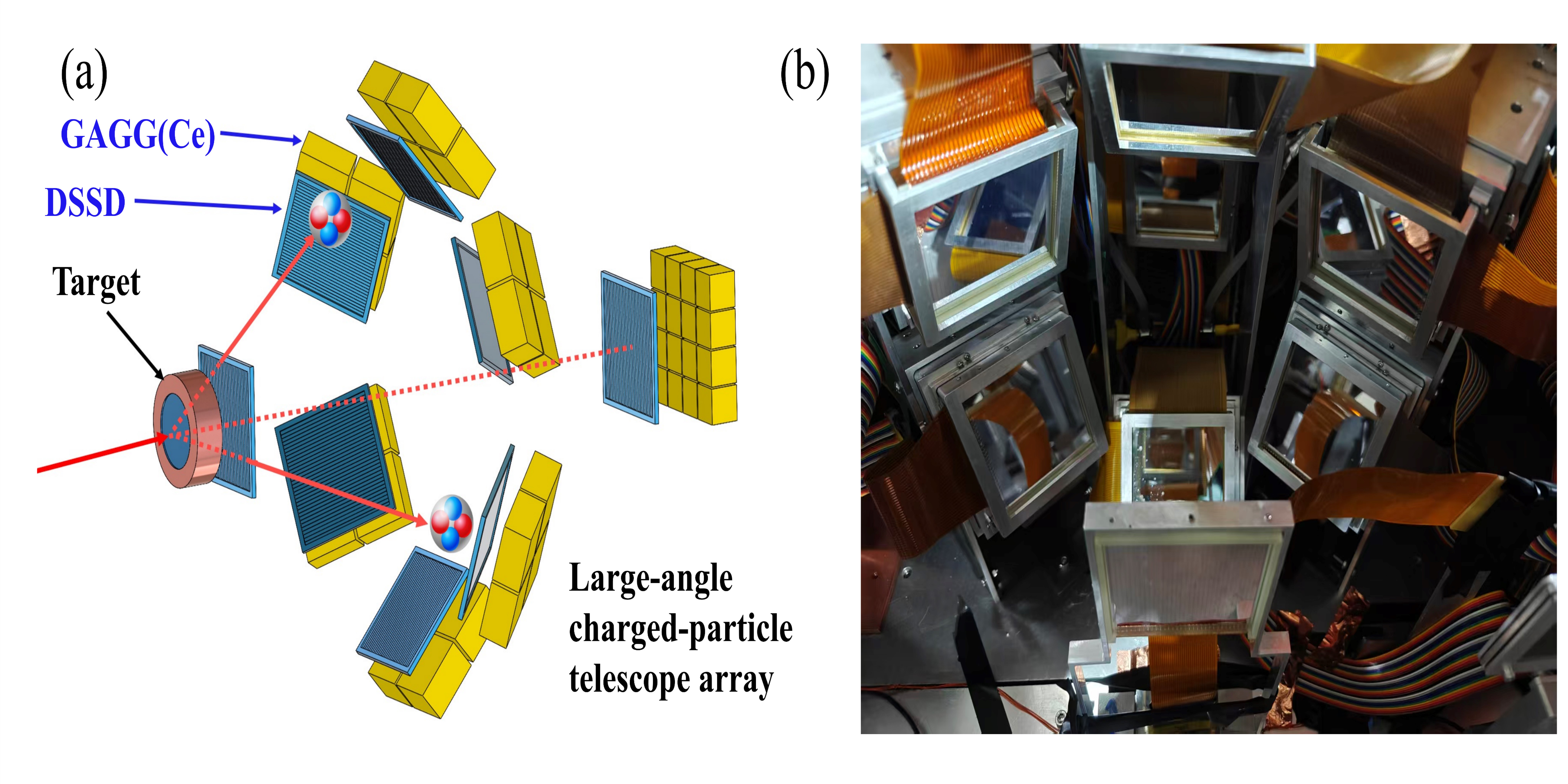}
  \caption{(a) Schematic diagram and (b) photograph of the experimental setup for the knockout reaction experiment at RIBLL1, Lanzhou.}\label{fig-ribll1_setup} 
\end{figure*}

The experimental setup is schematically shown in Fig. \ref{fig-ribll1_setup}. 
Six large-angle charged-particle telescope modules were installed downstream of the reaction target,  
covering a laboratory angular range from approximately 30° to 60°. 
Each telescope consisted of a DSSD followed by four GAGG(Ce) scintillator crystals, 
forming a $\mathrm{\Delta E - E}$ zero degree telescope configuration.  
The DSSDs had thicknesses of 500 $\mathrm{\mu m}$ or 1000 $\mathrm{\mu m}$, 
while the GAGG(Ce) scintillators were sufficiently thick to fully stop the reaction products of interest. 
The telescope array was designed to measure the energy and emission angles of light charged particles, 
in particular $\mathrm{\alpha}$ particles knocked out from the projectile.

Owing to the high channel density associated with the DSSDs in the telescope array, 
the GET electronics system was adopted for the readout of all DSSD channels of the large-angle charged-particle telescopes ($\sim$1000 channels in total).  
The beam energy was approximately 55 MeV/u, 
and based on kinematical simulations, the GET gain was set to 1 pC according to the expected energy deposition in the DSSDs. 
The ZAP interface boards with optimized DC-blocking capacitance values and the waveform-based energy extraction algorithms described above were directly applied in this experiment. 
The signals from GAGG(Ce) scintillators and other beamline detectors were read out using PIXIE-16 digital acquisition modules from XIA LLC\cite{hongyi-wu-general-purpose-2021,pixie16-web}, 
which were synchronized with GET using a common clock, 
enabling a hybrid DAQ system combining low-rate GET and high-rate electronics.

A critical aspect for the application of GET in physics measurements is its performance under different trigger rates. 
To quantify the event-loss behavior of the GET system, 
a dedicated trigger-rate test was performed using the MZ-TrigIO (MZTIO) module\cite{pu-digital-2024,mztio-web}. 
The MZTIO generated a periodic clock signal with a fixed frequency below 1 kHz, 
which was simultaneously fed as an external trigger to both the GET system and the PIXIE-16 modules. 
The PIXIE-16 system was used as a reference, since it can sustain trigger rates well above 100 kHz without event loss. 
After synchronizing the clocks of the two data acquisition systems, 
event losses in the GET system can be determined from the difference in the number of recorded events between GET and PIXIE-16.

\begin{figure}[htbp]
  \centering
  \includegraphics[width=0.7\textwidth]{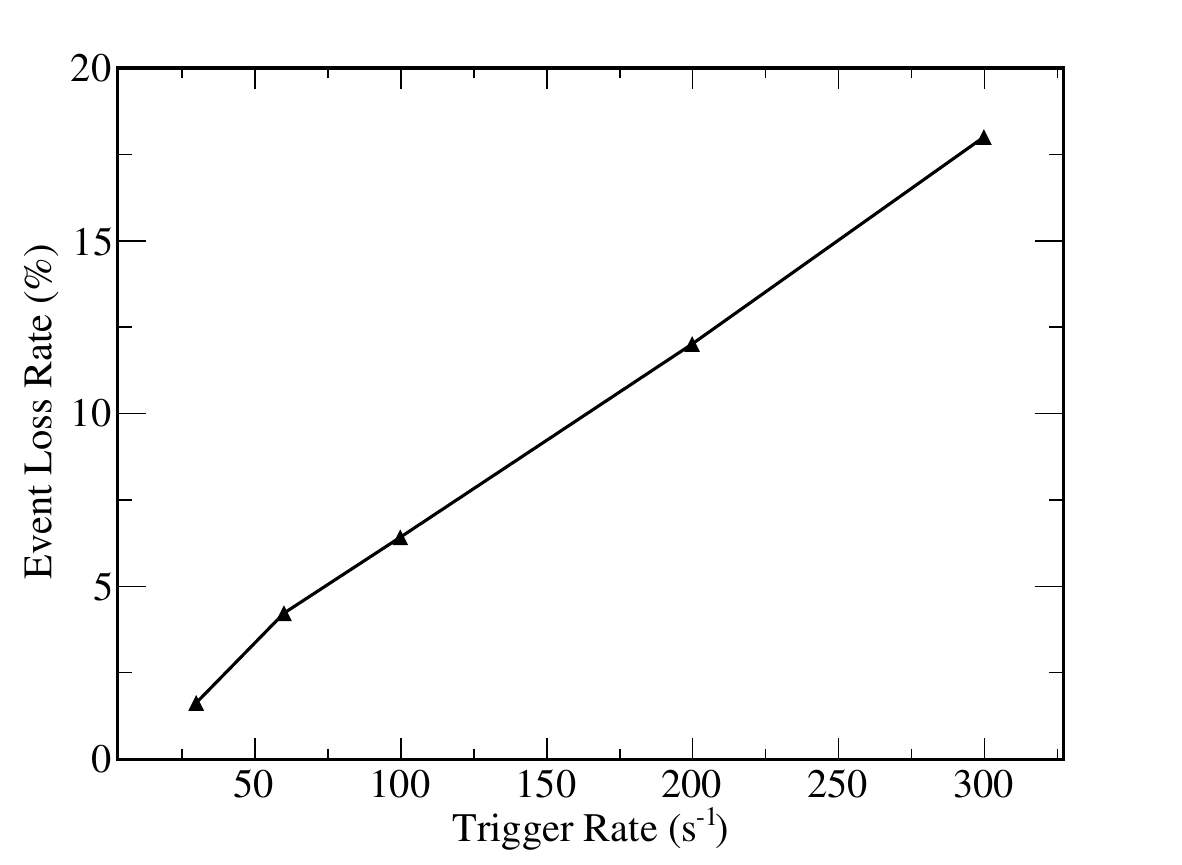}
  \caption{Event-loss ratio of the GET system as a function of trigger rate measured using an external clock trigger.}\label{fig-eventloss} 
\end{figure}

Fig. \ref{fig-eventloss} shows the measured event-loss ratio of the GET system as a function of trigger rate. 
A clear approximately linear increase in event loss with increasing trigger rate is observed, 
consistent with expectations for waveform-sampling electronics operating at relatively low trigger rates, as also reported in previous GET applications for CHIMERA and FARCOS\cite{deluca-test-2017, de-filippo-integration-2018}.
Therefore, during the physics run the GET system was operated in external-trigger mode with a carefully controlled trigger rate. 
The trigger logic combined signals from beamline detectors and GAGG(Ce) scintillators, 
and by adjusting the downscaling factors of the beamline trigger, the GET trigger rate was kept below approximately 220 Hz. 
Under these conditions, event losses were limited to below 10\%, 
and stable long-term operation was achieved without system instabilities or crashes.

\begin{figure}[htbp]
  \centering
  \includegraphics[width=0.7\textwidth]{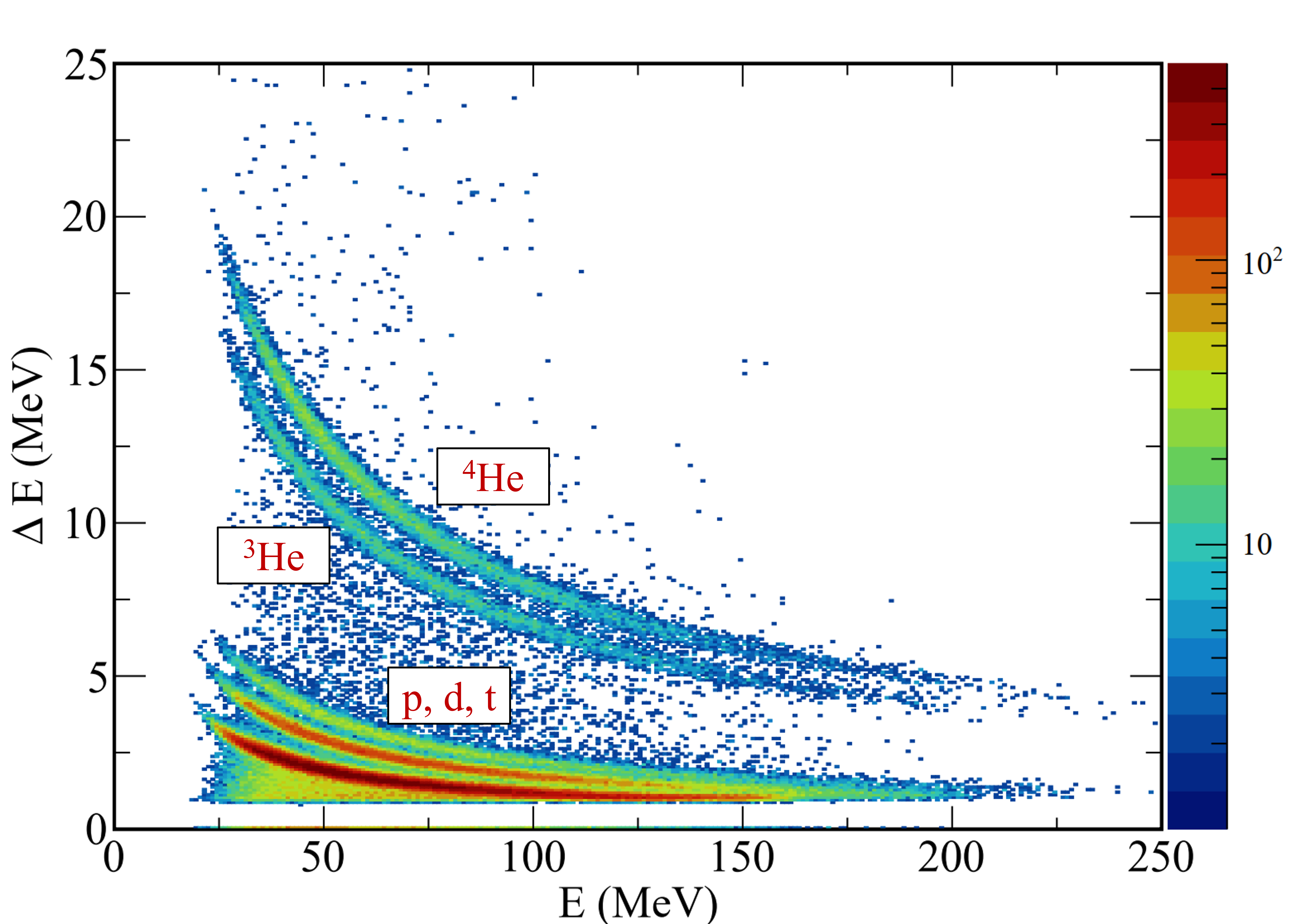}
  \caption{Particle-identification (PID) spectrum obtained from a large-angle telescope using the $\mathrm{\Delta E-E}$ method, where $\mathrm{\Delta E}$ and E correspond to the energy deposited in the DSSD and GAGG scintillators, respectively.}\label{fig-PID} 
\end{figure}

Fig. \ref{fig-PID} presents a representative particle-identification (PID) spectrum obtained with one of the large-angle telescopes during the breakup measurement of $^{11}$C, 
where $\mathrm{\Delta E}$ and $\mathrm{E}$ correspond to the energy deposited in the DSSD, read out by the GET system, and in the downstream GAGG(Ce) scintillators, respectively. 
Clear separation of protons, deuterons, tritons, as well as $^3$He and $^4$He particles is achieved, 
demonstrating the good energy resolution and stability of the GET-based DSSD readout in a realistic experimental environment. 
These results confirm the feasibility of employing the GET system for multi-channel DSSD readout in charged-particle telescope arrays 
and validate the hardware and software optimization methods proposed in this work.

\section{Summary}
In this work, we address the challenge of multi-channel readout for DSSDs in charged-particle detection by implementing and optimizing the GET system. 
An offline test platform based on a single DSSD, GET electronics, and custom-designed ZAP interface boards was first established to evaluate the feasibility and performance of the readout scheme.
The DC-blocking capacitors on ZAP1 were optimized by balancing signal amplitude, energy resolution, and the available dynamic range of the GET system.
In addition to hardware optimization, two waveform-based energy extraction algorithms were developed and compared with conventional direct peak-finding algorithm. 
Among them, template waveform fitting demonstrated superior robustness against electronic noise and provided the best overall energy resolution.
Using the optimized readout configuration, an energy resolution of $\sigma$ = 22 keV was achieved in the $\mathrm{\alpha}$-source test. 
The optimized GET-based DSSD readout system was further validated in a RIB physics experiment employing charged-particle telescopes, 
where clear particle identification of light charged particles was achieved. 
These results establish a scalable and versatile multi-channel DSSD readout solution based on GET electronics, with broad applicability to nuclear physics experiments.

\section*{CRediT authorship contribution statement}
\textbf{Jiawei BIAN}: Writing – original draft, Visualization, Software, 
Methodology, Investigation, Formal analysis, Data curation, Conceptualization. 
\textbf{Zaihong YANG}: Writing – review \& editing, Validation, Supervision,
Methodology, Investigation, Data curation, Conceptualization. 
\textbf{Qite LI}: Validation, Resources. 
\textbf{Zeyu DU}: Investigation, Validation, Resources.
\textbf{Xuan WANG}: Investigation, Validation, Resources.
\textbf{Ningtao ZHANG}: Validation, Resources.
\textbf{Pengjie LI}: Validation, Resources.
\textbf{Zhongtao SHEN}: Validation, Resources.
\textbf{Kaijie ZHOU}: Validation, Resources.
\textbf{Cheng WANG}: Validation, Resources.

\section*{Acknowledgements}
We would like to thank the HIRFL-RIBLL staff for their technical support.

This work was supported by the National KeyR\&D Program of China (Grants No. 2022YFA1605100, No. 2023YFE0101500), 
the National Natural Science Foundation of China (Grant No. 12275006) and the State Key Laboratory of Nuclear Physics and Technology, Peking University (Nos. NPT2024ZX01 and NPT2025ZX02).


\end{document}